\documentstyle[aps,amssymb,amsfonts,times,mathptm,12pt]{revtex}
\begin{document}
\small
\normalsize
\protect\newtheorem{principle}{Principle} %[section]
\protect\newtheorem{theo}[principle]{Theorem}
\protect\newtheorem{prop}[principle]{Proposition}
\protect\newtheorem{lem}[principle]{Lemma}
\protect\newtheorem{co}[principle]{Corollary}
\protect\newtheorem{de}[principle]{Definition}
\newtheorem{ex}[principle]{Example}
\newtheorem{rema}[principle]{Remark}
\newtheorem{rem}[principle]{Remark}
\newcounter{saveeqn}
\newcommand{\alpheqn}{\setcounter{saveeqn}{\value{equation}}%
\setcounter{equation}{0}%
\renewcommand{\theequation}{\mbox{\arabic{saveeqn}-\alph{equation}}}}
\newcommand{\reseteqn}{\setcounter{equation}{\value{saveeqn}}%
\renewcommand{\theequation}{\arabic{equation}}}
\renewcommand{\baselinestretch}{1}
\small
\normalsize
\title{Further results on the cross norm criterion
for separability}
\author{Oliver Rudolph \thanks{email: rudolph@fisicavolta.unipv.it}}
\address{Quantum Optics \& Information Group, Istituto Nazionale per la Fisica
della Materia \& Dipartimento \\ di Fisica "A.~Volta", Universita di
Pavia, via Bassi 6, I-27100 Pavia, Italy}
\maketitle
\begin{abstract}
\noindent In the present paper the cross norm criterion
for separability
of density matrices is studied.
In the first part of the paper  we determine the value
of the greatest cross
norm for Werner states, for isotropic states and for
Bell diagonal states. In the second part we
show that the greatest cross norm
criterion induces a novel
computable separability criterion for bipartite systems.
This new criterion is
a necessary but in general not a sufficient criterion for separability.
It is shown, however, that for all pure states,
for Bell diagonal states,
for Werner states in dimension $d=2$ and for
isotropic states in arbitrary dimensions the new criterion is
necessary and sufficient.
Moreover, it is shown that for Werner states in higher dimensions $d \geq
3$, the new criterion is only necessary.
\end{abstract}
\section{Introduction}
The greatest cross norm on the tensor product of the sets
of trace class operators on two (or more)
Hilbert spaces captures the concept
of entanglement in quantum theory in a mathematically natural way:
in \cite{Rudolph00} a separability criterion for mixed quantum
states was
proven using the greatest cross norm on the tensor product of
sets of trace class operators on finite dimensional Hilbert spaces.
It was shown that a density operator $\varrho$ is separable if and only if
the greatest cross norm of $\varrho$ is equal to 1.
In \cite{Rudolph00b} the value of the greatest cross
norm for pure states has been computed in terms of the Schmidt
coefficients of the state.
In the first part of this paper we determine the value
of the greatest cross norm for Werner states and for
isotropic states. We use methods
to compute entanglement measures
under symmetry recently discussed by Vollbrecht
and Werner \cite{VollbrechtW01} and by Terhal and Vollbrecht
\cite{TerhalV00}.
We also clarify the relationship of the greatest cross norm with
the robustness of entanglement and determine the value of the
greatest cross norm for Bell diagonal states.

In the second part of this paper we introduce and study a novel
necessary separability criterion for bipartite systems
induced by the greatest cross
norm criterion. We show that the new criterion completely
characterizes the separability properties of pure states,
Bell diagonal states,
isotropic states in arbitrary dimensions and Werner states in
dimension
$d=2$ while in dimension $d \geq 3$ some inseparable Werner states
satisfy
the criterion as well. Our results imply that the new criterion is
neither weaker nor stronger than the
Peres-Horodecki positive partial transpose (ppt)
criterion \cite{Peres96,Horodecki96b}. [We call a separability
criterion (A) weaker than a separability criterion (B)
if every state that
violates (A) also violates (B).]
Our results also show
that the new criterion is not weaker than both
the reduction criterion for separability \cite{Horodecki97},
and
the separability criterion introduced by Nielsen and Kempe
\cite{NielsenK01}. By the results of \cite{VollbrechtW02} this
also implies that our criterion is not weaker than the entropic
separability criteria based on the generalized R\'enyi and Tsallis
entropies.
Moreover, violating our criterion does not imply distillability.

This paper is organized as follows: In Section \ref{sec2}
we collect
some basic definitions and results. In Section \ref{2b} the
greatest cross norm is evaluated for operators of rank one.
In Section \ref{sec3} we proceed to compute the
value of the greatest cross norm for Werner states, in
Section \ref{sec4} for isotropic states and in Section
\ref{Bellgamma} for Bell diagonal states. In Section \ref{sec2.3}
we clarify the relation of the greatest cross norm with the
robustness of entanglement introduced in \cite{VidalT99}.
In Section \ref{secb} we introduce and study our computable
separability criterion.

Throughout this paper the set of trace class operators on some Hilbert
space ${\mathtt{H}}$ is denoted by ${\mathtt{T}}({\mathtt{H}})$, the set
of Hilbert-Schmidt operators on ${\mathtt{H}}$
by ${\mathtt{HS}}(\mathtt{H})$ and
the set of bounded operators on ${\mathtt{H}}$ by
${{\mathtt{B}}}({\mathtt{H}})$. A density operator is a positive trace
class operator with trace one. We use the Dirac bra/ket notation
throughout.

\section{Separability and the greatest cross norm}
\subsection{Preliminaries} \label{sec2}
\begin{de}
Let ${\mathtt{H}}_1$ and ${\mathtt{H}}_2$ be two Hilbert spaces of
arbitrary dimension. A density operator $\varrho$ on the tensor product
${\mathtt{H}}_1
\otimes {\mathtt{H}}_2$ is called \emph{separable} if there exist a
family $\left\{ \omega_{i} \right\}$ of positive real numbers, a family
$\left\{ \rho^{(1)}_i \right\}$ of density operators on
${\mathtt{H}}_1$ and a family $\left\{ \rho^{(2)}_i \right\}$ of
density operators
on ${\mathtt{H}}_2$ such that
\begin{equation} \label{e1}
\varrho = \sum_{i} \omega_{i} \rho^{(1)}_i \otimes \rho^{(2)}_i,
\end{equation}
where the sum converges in trace class norm.
A non-separable state is called {\em entangled}. \end{de}

The Schmidt decomposition is of central importance in the
characterization and quantification of entanglement associated
with pure states.
\begin{lem}
Let ${\mathtt{H}}_1$ and ${\mathtt{H}}_2$ be Hilbert spaces of
arbitrary dimension and let $\vert \psi \rangle
\in {\mathtt{H}}_1 \otimes
{\mathtt{H}}_2$.
Then there exist a family of non-negative real numbers $\{
p_i \}_i$ and orthonormal bases $\{ \vert a_i \rangle \}_i$ and
$\{ \vert b_i \rangle \}_i$ of
${\mathtt{H}}_1$ and ${\mathtt{H}}_2$ respectively such that
\[ \vert \psi \rangle = \sum_i \sqrt{p_i} \vert a_i \otimes b_i
\rangle. \] \label{Sch}
\end{lem}
The family of positive numbers $\{ p_i \}_i$ is called the family
of \emph{Schmidt coefficients} of $\vert \psi \rangle$.

Consider the spaces ${\mathtt{T}}({\mathtt{H}}_1)$ and ${\mathtt{T}}({\mathtt{H}}_2)$
of trace class operators on ${\mathtt{H}}_1$ and ${\mathtt{H}}_2$
respectively. Both spaces are Banach spaces when equipped with the trace
class norm $\Vert \cdot \Vert_1^{(1)}$ or $\Vert \cdot
\Vert_1^{(2)}$ respectively, see, e.g., Schatten \cite{Schatten70}. In
the sequel we shall drop the superscript and write $\Vert \cdot
\Vert_1$ for both norms, slightly abusing the notation; it will be
always clear from the context which norm is meant.
The algebraic tensor product ${\mathtt{T}}({\mathtt{H}}_1)
\otimes_{\rm alg} {\mathtt{T}}({\mathtt{H}}_2)$
of ${\mathtt{T}}({\mathtt{H}}_1)$
and ${\mathtt{T}}({\mathtt{H}}_2)$ is defined as the set of all finite
linear combinations of elementary tensors $u \otimes
{v}$, i.e., the set of all finite sums $\sum_{i=1}^n u_i \otimes
{v}_i$ where $u_i \in {\mathtt{T}}({\mathtt{H}}_1)$ and
${v}_i \in {\mathtt{T}}({\mathtt{H}}_2)$ for all $i$.
\begin{de}
A norm $\Vert \cdot \Vert$ on ${\mathtt{T}}({\mathtt{H}}_1)
\otimes_{\rm alg} {\mathtt{T}}({\mathtt{H}}_2)$ is called a
\emph{subcross norm} if $\Vert t_1 \otimes t_2 \Vert \leq
\Vert t_1 \Vert_1 \/ \Vert t_2 \Vert_1$ for all
$t_1 \in {\mathtt{T}}({\mathtt{H}}_1)$ and $t_2 \in
{\mathtt{T}}({\mathtt{H}}_2)$. It is called a {\em cross norm} if
$\Vert t_1 \otimes t_2 \Vert =
\Vert t_1 \Vert_1 \/ \Vert t_2 \Vert_1$ for all
$t_1 \in {\mathtt{T}}({\mathtt{H}}_1)$ and $t_2 \in
{\mathtt{T}}({\mathtt{H}}_2)$.
\end{de}
It is known that we can define a norm on ${\mathtt{T}}({\mathtt{H}}_1)
\otimes_{\rm alg} {\mathtt{T}}({\mathtt{H}}_2)$ by
\begin{equation} \label{cross}
\Vert t \Vert_\gamma := \inf \left\{ \sum_{i=1}^n
\left\Vert u_i \right\Vert_1 \, \left\Vert
{v}_i \right\Vert_1 \, \left\vert \, t = \sum_{i=1}^n u_i
\otimes {v}_i \right. \right\}, \end{equation} where $t \in
{\mathtt{T}}({\mathtt{H}}_1)
\otimes_{\rm alg} {\mathtt{T}}({\mathtt{H}}_2)$ and where the
infimum runs over
all finite decompositions of $t$ into elementary tensors
\cite{WeggeOlsen93}.

The norm $\Vert \cdot \Vert_\gamma$ defined in Equation (\ref{cross})
is born to be subcross and
can be shown to be cross (for a proof see, e.g., \cite{WeggeOlsen93}).
Moreover, $\Vert \cdot \Vert_\gamma$ majorizes any
subcross norm on ${\mathtt{T}}({\mathtt{H}}_1)
\otimes_{\rm alg} {\mathtt{T}}({\mathtt{H}}_2)$ and is therefore
often also referred to as the \emph{greatest cross norm}
on ${\mathtt{T}}({\mathtt{H}}_1)
\otimes_{\rm alg} {\mathtt{T}}({\mathtt{H}}_2)$.
The completion of
${\mathtt{T}}({\mathtt{H}}_1)
\otimes_{\rm alg} {\mathtt{T}}({\mathtt{H}}_2)$ with respect to
$\Vert \cdot \Vert_\gamma$ is denoted
by ${\mathtt{T}}({\mathtt{H}}_1)
\otimes_{\gamma} {\mathtt{T}}({\mathtt{H}}_2)$. In finite
dimensions we have
${\mathtt{T}}({\mathtt{H}}_1)
\otimes_{\gamma} {\mathtt{T}}({\mathtt{H}}_2) =
{\mathtt{T}}({\mathtt{H}}_1 \otimes
{\mathtt{H}}_2)$ \cite{WeggeOlsen93}.

In analogy we can also define a cross norm on ${\tt HS}({\mathtt{H}}_1)
\otimes_{\rm alg} {\tt HS}({\mathtt{H}}_2)$ by
\begin{equation}
\Vert t \Vert_g := \inf \left\{ \sum_{i=1}^n
\left\Vert u_i \right\Vert_2 \, \left\Vert
{v}_i \right\Vert_2 \, \left\vert \, t = \sum_{i=1}^n u_i
\otimes {v}_i \right. \right\}, \end{equation} where $t \in
{\tt HS}({\mathtt{H}}_1)
\otimes_{\rm alg} {\tt HS}({\mathtt{H}}_2)$ and where the infimum runs over
all finite decompositions of $t$ into elementary tensors. $\Vert \cdot
\Vert_2$ denotes the Hilbert-Schmidt norm.

In the following we are mainly interested in
the situation where both ${\mathtt{H}}_1$ and ${\tt
H}_2$ are finite dimensional, hence ${\mathtt{T}}({\mathtt{H}}_1) =
{\mathtt{B}}({\mathtt{H}}_1)$ and ${\mathtt{T}}({\mathtt{H}}_2) =
{\mathtt{B}}({\mathtt{H}}_2)$.

The following theorem demonstrates that $\Vert \cdot \Vert_\gamma$
captures the concept of entanglement
in quantum theory in a mathematically natural way. For a proof see \cite{Rudolph00}.
\begin{theo} \label{t1}
Let ${\mathtt{H}}_1$ and ${\mathtt{H}}_2$ be finite dimensional Hilbert
spaces
and $\varrho$ be a density operator on ${\mathtt{H}}_1
\otimes {\mathtt{H}}_2$. Then the following statements are equivalent:
\begin{itemize}
\item $\varrho$ is separable
\item
$\Vert \varrho \Vert_\gamma = 1$.
\end{itemize} \end{theo}
\subsection{Operators of rank one} \label{2b}
The following proposition is a slight generalization of a proposition that
 has been proven in \cite{Rudolph00b}.
It shows that on pure states $\Vert \cdot \Vert_\gamma$ can be
expressed by the Schmidt coefficients of the state.
We reproduce the proof here as the proof method is essential for
the results in Section \ref{secb}.
\begin{prop} \label{rank1}
Let ${\mathtt{H}}_1$ and ${\mathtt{H}}_2$ be finite dimensional
Hilbert spaces and let $\vert \psi \rangle, \vert \omega \rangle
\in {\mathtt{H}}_1 \otimes
{\mathtt{H}}_2$ be unit vectors and
$\vert \psi \rangle = \sum_i \sqrt{p_i}
\vert \phi_i \rangle
\otimes \vert \chi_i \rangle$ and
$\vert \omega \rangle = \sum_j \sqrt{q_j}
\vert \alpha_j \rangle
\otimes \vert \beta_j \rangle$
their Schmidt representations respectively. Here
$\{ \vert \phi_i \rangle \}_i$ and $\{ \vert \alpha_j \rangle \}_j$ are
orthonormal bases of ${\mathtt{H}}_1$ while
$\{ \vert \chi_i \rangle \}_i$ and $\{ \vert \beta_j \rangle \}_j$ are
orthonormal bases of ${\mathtt{H}}_2$. Moreover,
$p_i \geq 0$ and $q_j \geq 0$ and $\sum_i p_i = \sum_j q_j = 1$.
Let $S := \vert \psi \rangle \langle \omega \vert$. Then
\[ \Vert S \Vert_\gamma = \sum_{ij} \sqrt{p_i q_j}
= \left( \sum_i \sqrt{p_i} \right) \left( \sum_i
\sqrt{q_i} \right). \]
\end{prop}
\emph{Proof}: Without loss of generality we assume that
${\mathtt{H}}_1 = {\mathtt{H}}_2$ which can always be achieved
by possibly suitably enlarging one of the two Hilbert spaces.
Further, we
identify ${\mathtt{H}}_1 = {\mathtt{H}}_2$ with
${\mathbb{C}}^n$, where $n = \dim {\mathtt{H}}_1$, i.e., we fix
an orthonormal basis in ${\mathtt{H}}_1$ which we identify with the
canonical real basis in ${\mathbb{C}}^n$. With respect to this
canonical real basis in ${\mathbb{C}}^n$ we can define complex
conjugates of elements of ${\mathtt{H}}_1$ and the complex conjugate
as well as the transpose of a linear operator
acting on ${\mathtt{H}}_1$.
From the Schmidt decomposition it follows that
\begin{equation} S = \vert \psi \rangle \langle \omega \vert =
\sum_{ij}
\sqrt{p_i  q_j} \vert \phi_i \rangle \langle \alpha_j \vert \otimes
\vert \chi_i \rangle \langle \beta_j \vert. \label{0815}
\end{equation} From the definition of $\Vert \cdot \Vert_\gamma$ it
is thus obvious that $\Vert S \Vert_\gamma \leq \sum_{ij} \sqrt{p_i
q_j}$. Now consider the Hilbert space ${\mathtt{HS}}({\mathtt{H}}_1 \otimes
{\mathtt{H}}_2)$ of Hilbert-Schmidt operators
on ${\mathtt{H}}_1 \otimes {\mathtt{H}}_2$ equipped with the
Hilbert-Schmidt inner product $\langle f \vert g \rangle =
{\mathrm{tr}}(f^\dagger g)$. Equation (\ref{0815}) induces an
operator ${\mathfrak{A}}_S$ on ${\mathtt{HS}}({\mathtt{H}}_1 \otimes
{\mathtt{H}}_2)$ as follows.
Every element $\zeta$ in ${\mathtt{HS}}({\mathtt{H}}_1 \otimes
{\mathtt{H}}_2)$ can be written $\zeta =
\sum_k x_k \otimes y_k$ where $x_k$ and $y_k$ are trace class
operators on ${\mathtt{H}}_1$ and ${\mathtt{H}}_2$ respectively.
Then ${\mathfrak{A}}_S$ is defined on $\zeta$ as
${\mathfrak{A}}_S (\zeta) := \sum_{ijk} \sqrt{p_i q_j}
\langle \chi^*_i \vert x_k \vert \beta^*_j \rangle
\vert \phi_i \rangle \langle \alpha_j \vert \otimes y_k$ where
$\vert \chi_i^* \rangle$ and $\vert \beta_j^* \rangle$
denote, respectively, the complex conjugates of the vectors
$\vert \chi_i \rangle$ and $\vert \beta_j \rangle$
with respect to the canonical real basis in
${\mathbb{C}}^n$.
Proposition 11.1.8 in \cite{KadisonR86} implies that
${\mathfrak{A}}_S (\zeta)$ is independent of the
representation of $\zeta$. Consider a representation
$S = \sum_{i=1}^r u_i \otimes v_i$ of $S$ as sum over
simple tensors. Denote the transpose of $v_i$ by $v_i^T$.
Then the operator defined by
\begin{equation}
{\mathcal{A}}_S (\zeta) := \sum_{i,k=1}^r
{\mathrm{tr}}(v_i^T x_k) u_i \otimes y_k \label{huggies}
\end{equation} is equal to ${\mathfrak{A}}_S$ (by virtue of
Proposition 11.1.8 in \cite{KadisonR86}).
We denote the trace class norm on
${\mathtt{T}}({{\mathtt{HS}}({\mathtt{H}}_1 \otimes
{\mathtt{H}}_2)})$ by $\tau(\cdot)$. The operator
${\mathfrak{A}}_S$ is of trace class and the right hand side of
Equation (\ref{0815}) is the so-called polar representation
of ${\mathfrak{A}}_S$ which implies $\tau({\mathfrak{A}}_S) =
\sum_{ij} \sqrt{p_i q_j}$, see \cite{Schatten70}.
${\mathfrak{A}}_S$ admits also many
other representations ${\mathfrak{A}}_S \simeq \sum_i f_i \otimes
g_i$ with families of operators $\{ f_i \}$ and $\{ g_i \}$ acting
on ${\mathtt{H}}_1$ and ${\mathtt{H}}_2$ respectively.
It is known that \begin{equation} \label{tcn}
\tau({\mathfrak{A}}_S) = \inf \left\{
\sum_i \Vert f_i \Vert_2 \Vert g_i \Vert_2 \, \left\vert
{\mathfrak{A}}_S \simeq \sum_i f_i \otimes g_i \right.
\right\} \leq \Vert S \Vert_\gamma, \end{equation} where the latter
inequality follows from $\Vert z \Vert_2 \leq \Vert z \Vert_1$ and
from the fact that by construction
each decomposition of ${\mathfrak{A}}_S$
corresponds in an obvious one-to-one fashion
to a decomposition of $S$. For a proof of the first identity in
Equation (\ref{tcn}) see \cite{Schatten70}, page 42.
This proves the proposition. $\Box$
\begin{co} \label{au}
Let ${\mathtt{H}}_1$ and
${\mathtt{H}}_2$ be finite dimensional Hilbert spaces
and let $\rho$ be a density operator on ${\mathtt{H}}_1
\otimes {\mathtt{H}}_2$. Let
$\{ \vert \phi_i \rangle \}_i$ and $\{ \vert \alpha_j \rangle \}_j$ be
orthonormal bases of ${\mathtt{H}}_1$ and let
$\{ \vert \chi_i \rangle \}_i$ and $\{ \vert \beta_j \rangle \}_j$
be orthonormal bases of ${\mathtt{H}}_2$.
If $\rho = \sum_{ij}
a_{ij} \vert \phi_i \rangle \langle \alpha_j \vert \otimes
\vert \chi_i \rangle \langle \beta_j \vert$, then
$\Vert \rho \Vert_\gamma
= \sum_{ij} \vert a_{ij} \vert.$
\end{co}
Now consider the following expression
\begin{equation} \label{alpha}
\alpha : {\mathtt{T}}({\mathtt{H}}_1 \otimes {\mathtt{H}}_2) \to {\mathbb{R}},
\alpha(\sigma) := \inf \left\{ \sum_i \lambda_i \, \Vert
S_i \Vert_\gamma \, \right\vert \, \left.
\sigma = \sum_i \lambda_i S_i, \text{ where } \lambda_i \geq 0,
S_i \text{ of rank 1} \right\}
\end{equation} where the infimum is over all decompositions of
$\sigma$ into operators of rank 1. Obviously, $\Vert \sigma
\Vert_\gamma
\leq \alpha(\sigma)$ for all $\sigma \in {\mathtt{T}}({\mathtt{H}}_1)
\otimes_{\rm alg} {\mathtt{T}}({\mathtt{H}}_2)$.
We first show a little lemma
\begin{lem} \label{l8}
Let $\sigma \in {\mathtt{T}}({\mathtt{H}}_1)
\otimes_{\rm alg} {\mathtt{T}}({\mathtt{H}}_2)$, then
$\alpha(\sigma) = \Vert \sigma \Vert_\gamma$.
\end{lem}
\emph{Proof}: $\alpha$ is obviously a norm on ${\mathtt{T}}({\mathtt{H}}_1)
\otimes_{\rm alg} {\mathtt{T}}({\mathtt{H}}_2)$. Let $\sigma = \sigma_1
\otimes \sigma_2$ with $\sigma_1 \in {\mathtt{T}}({\mathtt{H}}_1)$ and
$\sigma_2 \in {\mathtt{T}}({\mathtt{H}}_2)$. Then let
$\sigma_1 = \sum_i \lambda_i^{(1)} S_i^{(1)}$ and
$\sigma_2 = \sum_j \lambda_j^{(2)} S_j^{(2)}$ be the polar
decompositions of $\sigma_1$ and $\sigma_2$ respectively
\cite{Schatten70}.
Then $S_i^{(1)}$ and $S_j^{(2)}$ are operators of rank 1 for all
$i,j$. Thus $\alpha(\sigma_1 \otimes \sigma_2) \leq
\sum_{ij} \left\vert \lambda_i^{(1)} \lambda_j^{(2)} \right\vert =
\Vert \sigma_1 \otimes \sigma_2 \Vert_1$.
This proves that $\alpha$ is a subcross
norm. As $\Vert \cdot \Vert_\gamma$ majorizes each subcross
seminorm we find that $\alpha(\sigma) \leq \Vert \sigma
\Vert_\gamma$ for all $\sigma \in {\mathtt{T}}({\mathtt{H}}_1 \otimes
{\mathtt{H}}_2)$. Hence $\alpha(\sigma) = \Vert \sigma \Vert_\gamma$.
$\Box$
\subsection{Werner states}
\label{sec3}
Let ${\mathtt{H}}$ be a finite dimensional Hilbert space and
let $d := \dim {\mathtt{H}} > 1$. Define
\[ {\mathbb{F}} := \sum_{i,j} \vert i \otimes j \rangle \langle
j \otimes i \vert \] where $(\vert i \rangle)$ is a orthonormal
basis of ${\mathtt{H}}$.
Werner states (first considered in \cite{Werner89})
are mixed quantum states in
${\mathtt{T}}({\mathtt{H}} \otimes {\mathtt{H}})$. They
can be parametrized by a real parameter $f$ with
$-1 \leq f \leq 1$ and are given by
\begin{equation}
\label{Werner} \varrho_f := \frac{1}{d^3 - d} \left(
(d-f) {\mathtt{1}} + (df-1) {\mathbb{F}} \right).
\end{equation} Note that ${\mathrm{tr}}(\varrho_f {\mathbb{F}}) =
f$. Let $\mathrm{G}$ be the group of all unitary operators on ${\mathtt{H}}
\otimes {\mathtt{H}}$ of the form $U \otimes U$ where $U$ is a unitary
on ${\mathtt{H}}$. Then a mixed quantum state is invariant under the
action of $\mathrm{G}$, i.e., $\varrho = V \varrho V^\dagger$ for all $V \in
\mathrm{G}$ if and only if $\varrho = \varrho_f$ for some $f$, see
\cite{Werner89}.
Define the \emph{twirling operator} ${\mathbf{P}}_G$ by
\[ {\mathbf{P}}_G(\sigma) \equiv \int dU (U \otimes U) \sigma (U^\dagger
\otimes
U^\dagger) \] where the integration is with respect to the Haar
measure of the unitary group on ${\mathtt{H}}$.

From the definitions of ${\mathbf{P}}_G$ and $\Vert \cdot \Vert_\gamma$
it readily follows that $\Vert {\mathbf{P}}_G (\sigma) \Vert_\gamma \leq
\Vert \sigma \Vert_\gamma$ for all $\sigma \in {\mathtt{T}}({\mathtt{H}}
\otimes {\mathtt{H}}).$

Let $M$ denote the set of
operators of rank 1 on ${\mathtt{H}} \otimes
{\mathtt{H}}$, and consider the expression
\begin{equation} \label{beta}
\beta(\varrho_f) := \inf \left\{ \sum_i \lambda_i \,
\Vert S_i \Vert_\gamma \, \right\vert \left. S_i \in M, \lambda_i
\geq 0,
\varrho_f = \sum_i \lambda_i {\mathbf{P}}_G (S_i) \right\}.
\end{equation}
\begin{lem} \label{betalemma}
Let $f \in [-1,1]$ and let $\varrho_f$ be the Werner state to $f$,
then $\beta(\varrho_f) = \Vert \varrho_f \Vert_\gamma$.
\end{lem}
\emph{Proof}:
Every admissible decomposition of $\varrho_f = \sum_i \lambda_i
S_i$ in Equation (\ref{alpha}) induces via $\varrho_f =
{\mathbf{P}}_G(\varrho_f) = \sum_i \lambda_i
{\mathbf{P}}_G(S_i)$
an admissible decomposition in Equation (\ref{beta}).
Hence $\beta(\varrho_f) \leq
\alpha(\varrho_f)$.
In turn for every decomposition $\varrho_f =
\sum_i \lambda_i {\mathbf{P}}_G(S_i)$ in Equation (\ref{beta})
we find $\Vert \varrho_f \Vert_\gamma \leq \sum_i \lambda_i \Vert
{\mathbf{P}}_G(S_i) \Vert_\gamma \leq \sum_i \lambda_i \Vert S_i
\Vert_\gamma$. Thus it follows that
also $\Vert \varrho_f \Vert_\gamma
\leq \beta(\varrho_f)$. $\Box$ \\

We are now ready to compute the greatest cross norm for Werner
states.
\begin{theo} \label{t9}
Let $\varrho_f$ be a Werner state, then
\begin{equation}
\Vert \varrho_f \Vert_\gamma = \left\{ \begin{array}{r@{\quad:\quad}l}
1 & \text{ for } \, 0 \leq f \leq 1 \\ 1 - f & \mbox{ for } -1 \leq f
< 0 \end{array} \right. .
\end{equation}
\end{theo}
\emph{Proof}:  Let $\varrho_f = \sum_i \lambda_i
{\mathbf{P}}_G (S_i)$ be an admissible decomposition in Equation
(\ref{beta}), then we write $S_i = \vert \varphi_i \rangle \langle
\psi_i \vert$ for all $i.$ We write the Schmidt decompositions of $\vert
\varphi_i \rangle$ and $\vert \psi_i \rangle$ as, respectively,
\begin{eqnarray*}
\vert \varphi_i \rangle & = & \sum_j \sqrt{p_j^{(i)}}
\left\vert a_j^{(i)} \otimes
b_j^{(i)} \right\rangle \\
\vert \psi_i \rangle
& = & \sum_k \sqrt{q_k^{(i)}} \left\vert d_k^{(i)} \otimes
e_k^{(i)} \right\rangle \end{eqnarray*} where
$\left(a^{(i)}_j \right)_j, \left(b^{(i)}_j \right)_j,
\left(d^{(i)}_k \right)_k$ and $\left(e^{(i)}_k \right)_k$
are orthonormal bases of ${\mathtt{H}}$ respectively for all $i$
and $\sum_j p^{(i)}_j
= \sum_k q^{(i)}_k = 1$ for all $i$.
The condition ${\mathrm{tr}}(\varrho_f
{\mathbb{F}}) = f$ reads
\begin{equation} f = \sum_{ijk}
\lambda_i \sqrt{p^{(i)}_j q^{(i)}_k} \left\langle
e^{(i)}_k  \right\vert \left. a^{(i)}_j \right\rangle \left\langle
d^{(i)}_k \right\vert \left. b^{(i)}_j \right\rangle.
\label{fcond} \end{equation}
Thus
\[ \beta(\varrho_f) = \inf \left\{ \sum_{ijk} \lambda_i
\sqrt{p^{(i)}_j q^{(i)}_k} \right\vert \left. \varrho_f =
\sum_i \lambda_i {\mathbf{P}}_G \left(
\vert \varphi_i \rangle \langle \psi_i \vert \right)
\right\} \] where the infimum is over
all decompositions of $\varrho_f$ of the form $\varrho_f =
\sum_i \lambda_i {\mathbf{P}}_G \left( \vert \varphi_i \rangle
\langle \psi_i \vert \right)$
and where $\left(p^{(i)}_j \right)_j$ and
$\left(q^{(i)}_k \right)_k$ are the Schmidt
coefficients of $\vert \varphi_i \rangle$ and
$\vert \psi_i \rangle$ respectively. Clearly $\beta(\varrho_f) \geq
1$. Now for $0 \leq f \leq 1$ choose $\lambda_i = \delta_{i1}$ and
$p^{(1)}_1 = q^{(1)}_1 = 1$ and $p^{(1)}_j = q^{(1)}_k = 0$ for
$j >  1$ and $k>1$. Moreover choose $\left\langle
e^{(1)}_1  \right\vert \left. a^{(1)}_1 \right\rangle =
\left\langle
d^{(1)}_1 \right\vert \left. b^{(1)}_1 \right\rangle = \sqrt{f}$,
then Equation (\ref{fcond}) is satisfied and $\sum_{ijk}
\vert \lambda_i \vert
\sqrt{p^{(i)}_j q^{(i)}_k} =1$ showing that the infimum is attained
$\Vert \varrho_f
\Vert_\gamma = \beta(\varrho_f) = 1$. In the case $-1 \leq f < 0$
we note that
\begin{eqnarray*}
1 - f & = & \sum_{ijk} \lambda_i \sqrt{p^{(i)}_j q^{(i)}_k}
\left( \left\langle d_k^{(i)} \right\vert \left. a_j^{(i)}
\right\rangle \left\langle e_k^{(i)}
\right\vert \left. b_j^{(i)} \right\rangle -
\left\langle e_k^{(i)} \right\vert \left. a_j^{(i)}
\right\rangle \left\langle d_k^{(i)} \right\vert \left.
b_j^{(i)} \right\rangle \right) \\ & \leq &
\sum_{ijk} \lambda_i \sqrt{p^{(i)}_j q^{(i)}_k}
\left\vert \left\langle d_k^{(i)} \right\vert \left.
a_j^{(i)} \right\rangle \left\langle e_k^{(i)}
\right\vert \left. b_j^{(i)} \right\rangle - \left\langle e_k^{(i)}
\right\vert \left. a_j^{(i)}
\right\rangle \left\langle d_k^{(i)} \right\vert \left.
b_j^{(i)} \right\rangle \right\vert \\
& \leq & \sum_{ijk} \lambda_i \sqrt{p^{(i)}_j q^{(i)}_k}.
\end{eqnarray*} The last inequality follows readily by considering
the Schmidt decomposition of an unnormalized vector of the form
$\vert d \otimes e - e \otimes d \rangle$.
Thus in general $\beta(\varrho_f) \geq 1 -f$. Now
choose $\lambda_i = \delta_{i1}$ again and
\[ \vert \varphi_1 \rangle = \vert \psi_1 \rangle \equiv
\sqrt{p^{(1)}_1} \left\vert a^{(1)}_1 \otimes b^{(1)}_1 \right\rangle -
\sqrt{p^{(1)}_2}
\left\vert b^{(1)}_1 \otimes a^{(1)}_1 \right\rangle \]
where $p_1^{(1)} =
1 - p_2^{(1)} = \frac{1}{2} - \frac{1}{2} \sqrt{1 - f^2}$
and
where $\left\vert a^{(1)}_1 \right\rangle$ and $\left\vert
b^{(1)}_1 \right\rangle$
satisfy $\left\langle b^{(1)}_1 \right\vert \left. a^{(1)}_1
\right\rangle = 0$.
For this choice Equation (\ref{fcond}) is satisfied and
we have $1 = \sum_{ij}
\lambda_i \sqrt{p^{(i)}_j p^{(i)}_j}$ and
$-f = \sum_{ijk \atop j \neq k}
\lambda_i \sqrt{p^{(i)}_j p^{(i)}_k}$. Thus
$\Vert \varrho_f \Vert_\gamma = \beta(\varrho_f) = 1-f$.
$\Box$
\subsection{Isotropic states}
\label{sec4}
Again let ${\mathtt{H}}$ be a finite dimensional Hilbert space with
dimension $d := \dim {\mathtt{H}} > 1$.
Consider the group $\widehat{G}$ of a local
unitary transformations on ${\mathtt{H}} \otimes {\mathtt{H}}$ of the form
$U \otimes \bar{U}$ where $U$ is a unitary on ${\mathtt{H}}$ and
$\bar{U}$ denotes the complex conjugate of $U$ with respect to an
arbitrary but fixed orthonormal basis in ${\mathtt{H}}$. The set of
states invariant under all elements of $\widehat{G}$ are the
so-called \emph{isotropic states}, see, e.g.,
\cite{Horodecki97,VollbrechtW01,TerhalV00}.
The isotropic states can be parametrized by a
positive real parameter $F \in [0,1]$ and are given by
\begin{equation} \label{isotropic}
\varrho_F \equiv \frac{1-F}{d^2-1} \left( {\mathtt{1}} -
\left\vert \Psi^+ \right\rangle \left\langle \Psi^+ \right\vert
\right) + F \left\vert \Psi^+ \right\rangle \left\langle
\Psi^+ \right\vert. \end{equation}
Here $\left\vert \Psi^+ \right\rangle \equiv \frac{1}{\sqrt{d}}
\sum_{i=1}^d \vert i \otimes i \rangle$ and $\left( \vert i \rangle
\right)_i$ is an arbitrary orthonormal basis in ${\mathtt{H}}$.
We define \[ \widehat{\mathbb{F}} := d \left\vert \Psi^+
\right\rangle \left\langle \Psi^+ \right\vert =
\sum_{ij} \vert i \otimes i \rangle \langle j \otimes j \vert. \]
Then ${\mathrm{tr}}(\varrho_F \widehat{\mathbb{F}}) = dF$.
We proceed in analogy to Section \ref{sec3} and define the twirling
operator $\widehat{\mathbf{P}}_{\widehat{G}}$ for $\widehat{G}$ by
\[ \widehat{\mathbf{P}}_{\widehat{G}}(\sigma) :=
\int dU (U \otimes \bar{U}) \sigma (U^\dagger
\otimes \bar{U}^\dagger) \] where the integration is again with
respect to the Haar measure of the unitary group on ${\mathtt{H}}$.
Let $M$ denote the set of operators of rank 1 on ${\mathtt{H}} \otimes
{\mathtt{H}}$, then consider the expression
\begin{equation} \label{betahat}
\widehat{\beta}(\varrho_F) :=
\inf \left\{ \sum_i \lambda_i
\Vert S_i \Vert_\gamma \, \right\vert \left. S_i \in M, \lambda_i
\geq 0,
\varrho_F = \sum_i \lambda_i \widehat{{\mathbf{P}}}_{\widehat{G}}
(S_i) \right\}.
\end{equation}
\begin{lem}
Let $F \in [0,1]$ and let $\varrho_F$ be the isotropic state to $F$,
then $\widehat{\beta}(\varrho_F)
= \Vert \varrho_F \Vert_\gamma$.
\end{lem}
\emph{Proof}: Analogous to the proof of Lemma \ref{betalemma}.
$\Box$
\begin{theo} \label{t10}
Let $F \in [0,1]$ and $\varrho_F$ be the isotropic state to $F$, then
\begin{equation}
\Vert \varrho_F \Vert_\gamma = \left\{ \begin{array}{r@{\quad:\quad}l}
1 & \text{ for } \, \, 0 \leq F \leq \frac{1}{d} \\
dF & \mbox{ for } \, \frac{1}{d} < F
\leq 1 \end{array} \right. .
\end{equation}
\end{theo}
\emph{Proof}: The proof proceeds in analogy to the proof of
Theorem \ref{t9}.
Let $\varrho_F = \sum_i \lambda_i
\widehat{\mathbf{P}}_{\widehat{G}} (S_i)$
be an admissible decomposition in Equation
(\ref{betahat}),
then we write $S_i = \vert \varphi_i \rangle \langle
\psi_i \vert$ for all $i.$
We write the Schmidt decompositions of $\vert
\varphi_i \rangle$ and $\vert \psi_i \rangle$ as
$\vert \varphi_i \rangle  =  \sum_j \sqrt{p_j^{(i)}}
\left\vert a_j^{(i)} \otimes
b_j^{(i)} \right\rangle$ and
$\vert \psi_i \rangle
 =  \sum_k \sqrt{q_k^{(i)}} \left\vert d_k^{(i)} \otimes
e_k^{(i)} \right\rangle $
where
$\left(a^{(i)}_j \right)_j, \left(b^{(i)}_j \right)_j,
\left(d^{(i)}_k \right)_k$ and $\left(e^{(i)}_k \right)_k$
are orthonormal bases of ${\mathtt{H}}$ respectively for all $i$
and $\sum_j p^{(i)}_j
= \sum_k q^{(i)}_k = 1$ for all $i$.
The condition ${\mathrm{tr}}(\varrho_F
\widehat{\mathbb{F}}) = dF$ reads
\begin{equation} dF = \sum_{ijk} \lambda_i \sqrt{p^{(i)}_j q^{(i)}_k}
\left\langle
{a^{(i)}_j}^*  \right\vert \left. b^{(i)}_j
\right\rangle \left\langle
d^{(i)}_k \right\vert \left. {e^{(i)}_k}^* \right\rangle
\label{fcond3} \end{equation} where $\left\vert {a^{(i)}_j}^*
\right\rangle$
and $\left\vert {e^{(i)}_k}^* \right\rangle$ denote the complex conjugates of
$\left\vert {a^{(i)}_j} \right\rangle$ and $\left\vert {e^{(i)}_k}
\right\rangle$ respectively. Thus
\[ \widehat{\beta}(\varrho_F) =
\inf \left\{ \sum_{ijk} \lambda_i
\sqrt{p^{(i)}_j q^{(i)}_k} \right\vert \left. \varrho_F =
\sum_i \lambda_i \widehat{\mathbf{P}}_{\widehat{G}} \left(
\vert \varphi_i \rangle \langle \psi_i \vert \right)
\right\} \] where the infimum is over
all decompositions of $\varrho_f$ of the form $\varrho_f =
\sum_i \lambda_i {\mathbf{P}}_{\widehat{G}} \left( \vert \varphi_i \rangle
\langle \psi_i \vert \right)$
and where $\left(p^{(i)}_j \right)_j$ and
$\left(q^{(i)}_k \right)_k$ are the Schmidt
coefficients of $\vert \varphi_i \rangle$ and
$\vert \psi_i \rangle$ respectively.
From (\ref{fcond3}) it follows immediately that $\beta(\varrho_F) \geq
dF$.

For $dF \geq 1$, consider a state
$\vert \psi \rangle$ of the form
$\vert \psi \rangle = \sum_i \sqrt{\mu_i} \vert e_i \otimes e_i \rangle$
where $\{ e_i \}_i$ is an orthonormal basis  of $\mathtt{H}$ and
where $\left(\sum_i \sqrt{\mu_i} \right)^2 = dF$.
It has been shown in \cite{TerhalV00}
that $\varrho_F = \widehat{\mathbf{P}}_{\widehat{G}}(\vert \psi \rangle
\langle \psi \vert)$.
Now Proposition \ref{rank1} implies that $\Vert \varrho_F \Vert_\gamma
= \beta(\varrho_F) = dF$.

For $0 \leq dF < 1$, consider two states
$\vert a \rangle$ and $\vert b \rangle$ in $\mathtt{H}$ with
$\langle a^* \vert b \rangle = \sqrt{dF}$. Again, it has been shown
in \cite{TerhalV00} that $\varrho_F =
\widehat{\mathbf{P}}_{\widehat{G}}(\vert a \otimes b \rangle \langle
a \otimes b \vert).$ As by Theorem \ref{t1} we have
$\beta(\varrho_F) \geq 1$,
this proves $\Vert \varrho_F \Vert_\gamma = \beta(\varrho_F) = 1$.
$\Box$

\subsection{Bell diagonal states} \label{Bellgamma}
Consider ${\mathbb{C}}^2$ and let $\{
\vert 1 \rangle,
\vert 2 \rangle \}$ be an orthonormal basis of ${\mathbb{C}}^2$.
Then the Bell basis of
${\mathbb{C}}^2 \otimes {\mathbb{C}}^2$ is given by
\begin{eqnarray*} \vert \Psi_0 \rangle \equiv
\frac{1}{\sqrt{2}} \vert 11 \rangle
+ \vert  22 \rangle, & \, \, &
\vert \Psi_1 \rangle \equiv \frac{i}{\sqrt{2}} \vert 12 \rangle
+ \vert  21 \rangle \\ \vert \Psi_2 \rangle \equiv
\frac{1}{\sqrt{2}} \vert 21 \rangle
- \vert  12 \rangle, & \, \, & \vert \Psi_3 \rangle \equiv
\frac{i}{\sqrt{2}} \vert 11 \rangle
- \vert  22 \rangle. \end{eqnarray*}
Bell diagonal states are the density operators on ${\mathbb{C}}^2
\otimes
{\mathbb{C}}^2$ which are diagonal in the Bell basis
\[ \varrho =\sum_{i=0}^3 \lambda_i \vert \Psi_i \rangle \langle
\Psi_i \vert. \]
Bell diagonal states are known to be separable if and only if $\lambda_i
\leq \frac{1}{2}$ for all $i$, \cite{Horodecki96,Bennett96}.
\begin{theo} \label{gammabell}
Let $\varrho \in {\mathtt{T}}({\mathbb{C}}^2 \otimes
{\mathbb{C}}^2)$ be a Bell diagonal state, i.e.,
$\varrho = \sum_{i=0}^3 \lambda_i \vert \Psi_i \rangle \langle
\Psi_i \vert$ with
$\lambda_i \geq 0$ for all $i$. Then
\begin{equation}
\Vert \varrho \Vert_\gamma =
\left\{ \begin{array}{r@{\quad:\quad}l}
2 \max_i \lambda_i & \text{ for } \, \, \max_i \lambda_i > \frac{1}{2}
\\ 1 & \mbox{ for } \, \, \max_i \lambda_i \leq \frac{1}{2}
\end{array} \right. .
\end{equation}
\end{theo}
\emph{First part of the proof}: First consider the case $\max_i
\lambda_i \leq \frac{1}{2}$. In this case there is an explicit
decomposition of $\varrho$ as a mixture of eight unentangled pure
states (see \cite{Bennett96} for details).
Thus $\Vert \varrho \Vert_\gamma = 1$.
Now consider the case that $\max_i \lambda_i > \frac{1}{2}$.
In this case there exists an explicit
decomposition of $\varrho$ as an equal probability mixture
of eight entangled pure states, each of which has
$\left\{ {\frac{1}{2} + \frac{1}{2}
\sqrt{2 \max_i \lambda_i - 4 (\max_i \lambda_i)^2}},
{\frac{1}{2} - \frac{1}{2}
\sqrt{2 \max_i \lambda_i - 4 (\max_i \lambda_i)^2}}
\right\}$ as its Schmidt coefficients (again, see \cite{Bennett96} for
details). From the subadditivity of $\Vert \cdot \Vert_\gamma$
and Proposition \ref{rank1} it
follows readily that $\Vert \varrho \Vert_\gamma \leq
2 \max_i \lambda_i.$ We postpone the proof for the remaining
inequality
$\Vert \varrho
\Vert_\gamma \geq 2 \max_i \lambda_i$ until Section \ref{Bell}. $\Box$

\subsection{Relationship with the robustness of entanglement}
\label{sec2.3}
Denote the set of Hermitean trace class operators on a Hilbert
space ${\mathtt{H}}$ by
${\mathtt{T}}^{{\mathrm{h}}}({\mathtt{H}})$.
A norm closely related to $\Vert \cdot \Vert_\gamma$
can be defined on ${\mathtt{T}}^{{\mathrm{h}}}({\mathtt{H}}_1)
\otimes_{\mathrm{alg}} {\mathtt{T}}^{{\mathrm{h}}}({\mathtt{H}}_2)$ by
\begin{equation} \label{base}
\Vert t \Vert_S := \inf \left\{ \sum_{i=1}^n
\left\Vert u_i \right\Vert_1 \, \left\Vert
{v}_i \right\Vert_1 \, \left\vert \, t = \sum_{i=1}^n u_i
\otimes {v}_i \right. \right\}
\end{equation} where $t \in
{\mathtt{T}}^{{\mathrm{h}}}({\mathtt{H}}_1)
\otimes_{\rm alg} {\mathtt{T}}^{{\mathrm{h}}}({\mathtt{H}}_2)$ and where the
infimum runs over
all finite decompositions of $t$ into elementary \emph{Hermitean}
tensors. From the definitions of $\Vert \cdot \Vert_\gamma$ and
$\Vert \cdot \Vert_S$ it is obvious that in general
$\Vert t \Vert_\gamma \leq \Vert t \Vert_S$
for all Hermitean trace class operators $t$. For a density operator
$\sigma$ it is also obvious
that $\Vert \sigma \Vert_S = 1$ if and only if $\sigma$ is
separable. Clearly, $\Vert \cdot \Vert_S$ is the greatest cross
norm on ${\mathtt{T}}^{{\mathrm{h}}}({\mathtt{H}}_1)
\otimes_{\mathrm{alg}}
{\mathtt{T}}^{{\mathrm{h}}}({\mathtt{H}}_2)$.

\begin{lem}
Let $\mathtt{H}$ be a finite dimensional Hilbert space and let
$\sigma$ be a Hermitean operator on ${\mathtt{H}} \otimes {\mathtt{H}}$, then
\[ \Vert \sigma \Vert_S = \kappa(\sigma) :=
\inf \left\{ a_+ + a_- \left\vert \sigma
= a_+ \varrho_+ - a_- \varrho_-, a_\pm \geq 0, \varrho_\pm
{\text{ separable density operators}} \right. \right\}. \]
\end{lem}
\textbf{Proof}: Obviously, for
every $\sigma$ there are $a_\pm \geq 0$ and separable density
operators $\varrho_\pm$ such that $\sigma
= a_+ \varrho_+ - a_- \varrho_-$. [It is always possible to
write $\sigma$ as a sum of Hermitean simple tensors $\sigma =
\sum_i x_i \otimes y_i$; to get the desired decomposition just
decompose all $x_i$ and $y_i$ into their positive and negative parts
and rearrange terms]. The inequality
$\Vert \sigma \Vert_S \leq \kappa(\sigma)$ is obvious. If
$\sigma = \sigma_1 \otimes \sigma_2$, then $\kappa(\sigma) \leq
\Vert \sigma_1 \Vert_1 \, \Vert \sigma_2 \Vert_1$. Thus $\kappa$
is a subcross norm and thus $\kappa(\sigma) \leq \Vert \sigma
\Vert_S$ for all Hermitean $\sigma$. $\Box$ \\

For a density operator $\sigma$
the quantity $E_R(\sigma) \equiv \frac{1}{2} \left(
\Vert \sigma \Vert_S -1 \right)$ is called \emph{robustness of
entanglement} \cite{VidalT99}, see also \cite{VidalW01}.
The robustness of entanglement has
the physical meaning of the minimal amount of separable noise that
destroys the entanglement of a given state.

\begin{prop} \label{p14}
Let $\mathtt{H}$ be a finite dimensional Hilbert space. Then the
robustness of entanglement and the greatest cross norm on
${\mathtt{T}}({\mathtt{H}} \otimes {\mathtt{H}})$ are related by
\begin{equation} \label{robust} E_R(\sigma) \geq
\Vert \sigma \Vert_\gamma - 1 \end{equation}
where $\sigma$ is a positive trace class operator with trace one.
\end{prop}
\emph{Proof}: The analogue of Lemma \ref{l8} holds for $\Vert \cdot
\Vert_S$. Let $\vert \psi \rangle \in {\mathtt{H}} \otimes
{\mathtt{H}}$ and let $P_\psi =
\vert \psi \rangle \langle \psi \vert.$
In \cite{VidalT99} it has been shown that
$\Vert P_\psi \Vert_S = 2 \left(
\sum_i \sqrt{p_i} \right)^2 - 1$.
Thus Proposition
\ref{rank1} implies that $\Vert P_\psi \Vert_S = 2
\Vert P_\psi \Vert_\gamma -1$. Therefore Lemma \ref{l8} and the analogue statement
for $\Vert \cdot \Vert_S$ imply that $\Vert \sigma \Vert_S \geq
2 \Vert \sigma \Vert_\gamma -1$ for all positive
Hermitean $\sigma$ with trace one.
This proves the proposition. $\Box$ \\

For projection operators we have equality in Equation
(\ref{robust}).
Moreover, in \cite{VidalW01} Vidal and Werner computed the robustness of
entanglement for density operators with symmetry.
The results of Vidal and Werner
show that for Werner and isotropic states there is also an
equality in Equation (\ref{robust}). However, a proof of whether or
not equality holds in (\ref{robust}) in general has not been found
by this author.
\section{A computable separability criterion}
\label{secb}
\subsection{Formulation of the criterion}
Every finite dimensional Hilbert space ${\mathtt{H}}$ is
isomorphic to ${\mathbb{C}}^n$, with $n = \dim({\mathtt{H}})$.
This corresponds to identifying a fixed
orthonormal basis in ${\mathtt{H}}$ with the canonical real basis
in ${\mathbb{C}}^n$. In ${\mathbb{C}}^n$
there is a notion of complex conjugation. We
denote the complex conjugate of $\vert \psi \rangle \in {\mathbb{C}}^n$
by $\vert \psi^* \rangle$.
\begin{prop} \label{prop3}
Let ${\mathtt{K}}_1 \simeq {\mathbb{C}}^n$ and ${\mathtt{K}}_2
\simeq {\mathbb{C}}^m$ be finite dimensional
Hilbert spaces. There is a one-to-one
correspondence between states
$\vert \psi \rangle \in {\mathtt{K}}_1 \otimes {\mathtt{K}}_2$ and
Hilbert-Schmidt operators $A : {\mathtt{K}}_2 \to {\mathtt{K}}_1$
according to the rule: let $\vert \psi \rangle =
\sum_{ij} c_{ij} \vert a_i \rangle
\otimes
\vert b_j \rangle$ be a decomposition of $\vert \psi \rangle$
in terms of orthonormal bases $\{ \vert a_i \rangle
\}$ and $\{ \vert b_j \rangle
\}$ of ${\mathtt{K}}_1$ and ${\mathtt{K}}_2$ respectively. Then
$A(\psi)$ is given by $A(\psi) =
\sum_{ij} c_{ij} \vert a_i \rangle \langle b_j^* \vert$.
Conversely, if $A =
\sum_{ij} c_{ij} \vert a_i \rangle \langle b_j \vert$ for some
orthonormal bases $\{ \vert a_i \rangle
\}$ and $\{ \vert b_j \rangle
\}$ of ${\mathtt{K}}_1$ and ${\mathtt{K}}_2$ respectively, then
$\vert \psi_A \rangle = \sum_{ij} c_{ij} \vert a_i \rangle
\otimes \vert b_j^* \rangle$. \end{prop}
\emph{Proof}: We only need to show that
$A(\psi)$ is well-defined and independent of the decomposition
of $\vert \psi \rangle$ and similarly that $\vert \psi_A \rangle$ is
independent of the representation of $A$ chosen.
But this follows immediately
from, e.g., Proposition 11.1.8 in \cite{KadisonR86}.
$\Box$
\begin{co}
Let ${\mathtt{K}}_1 \simeq {\mathbb{C}}^n$ and ${\mathtt{K}}_2
\simeq {\mathbb{C}}^m$ be finite dimensional Hilbert spaces. The
one-to-one correspondence between pure states $\vert \psi \rangle
\in {\mathtt{K}}_1 \otimes {\mathtt{K}}_2$ and Hilbert-Schmidt
operators $A : {\mathtt{K}}_2 \to {\mathtt{K}}_1$ from Proposition
\ref{prop3} is isometric, i.e., $\langle A(\psi_1) \vert A(\psi_2)
\rangle_{HS} = \langle \psi_1 \vert \psi_2 \rangle$ and $\langle
\psi_A \vert \psi_B \rangle = \langle A \vert B \rangle_{HS}.$
\end{co} \emph{Proof}: Denote the canonical real bases of
${\mathtt{K}}_1$ and ${\mathtt{K}}_2$ by $\{ \vert e_i \rangle
\}_i$ and $\{ \vert f_j \rangle \}_j$ respectively. Let $\vert
\psi_1 \rangle = \sum_{ij} c_{ij} \vert e_i \rangle \otimes \vert
f_j \rangle$ and $\vert \psi_2 \rangle = \sum_{pq} d_{pq} \vert
e_p \rangle \otimes \vert f_q \rangle$ the decompositions of
$\vert \psi_1 \rangle \in {\mathtt{K}}_1$ and $\vert \psi_2
\rangle \in {\mathtt{K}}_2$ in terms of these bases. Then
$A(\psi_1) = \sum_{ij} c_{ij} \vert e_i \rangle \langle f^*_j
\vert = \sum_{ij} c_{ij} \vert e_i \rangle \langle f_j \vert$ and
$A(\psi_2) = \sum_{pq} d_{pq} \vert e_p \rangle \langle f^*_q
\vert = \sum_{pq} d_{pq} \vert e_p \rangle \langle f_q \vert$.
Therefore $\langle A(\psi_1) \vert A(\psi_2) \rangle_{HS} =
{\mathrm{tr}}(A^\dagger(\psi_1) A(\psi_2))
= \sum_{ijpq} c_{ij}^* d_{pq} \langle f_q \vert
f_j \rangle \, \langle e_i \vert e_p \rangle =
\sum_{ijpq} c_{ij}^* d_{pq} \langle f_j \vert
f_q \rangle \, \langle e_i \vert e_p \rangle
= \langle \psi_1 \vert \psi_2 \rangle.$
This proves the corollary. $\Box$ \\

To derive the next theorem we use Proposition
\ref{prop3} in the case that ${\mathtt{K}}_1$ and ${\mathtt{K}}_2$
are the spaces of Hilbert-Schmidt operators on some other Hilbert
spaces ${\mathtt{H}}_1$ and ${\mathtt{H}}_2$ respectively, i.e.,
${\mathtt{K}}_1 = {\mathtt{HS}}({\mathtt{H}}_1)$ and
${\mathtt{K}}_2 = {\mathtt{HS}}({\mathtt{H}}_2)$.
\begin{theo}
Let ${\mathtt{H}}_1$ and ${\mathtt{H}}_2$ be finite dimensional
Hilbert spaces and let ${\mathtt{K}}_1 = {\mathtt{HS}}({\mathtt{H}}_1)
\simeq {\mathbb{C}}^n$ and ${\mathtt{K}}_2 = {\mathtt{HS}}({\mathtt{H}}_2)
\simeq {\mathbb{C}}^m$ be the spaces of Hilbert-Schmidt operators on
${\mathtt{H}}_1$ and ${\mathtt{H}}_2$ respectively.
Then there exists a one-to-one correspondence between
Hilbert-Schmidt operators $T \in {\mathtt{HS}}({\mathtt{H}}_1
\otimes
{\mathtt{H}}_2)$ and Hilbert-Schmidt operators ${\mathfrak{A}}(T) :
{\mathtt{HS}}(\mathtt{H}_2) \to {\mathtt{HS}}({\mathtt{H}}_1)$
analogous to the correspondence in Proposition \ref{prop3}.
\label{prop4} \end{theo}
\emph{Proof}: It is well-known that
${\mathtt{HS}}({\mathtt{H}})$ furnished with the
Hilbert-Schmidt inner product $\langle A \vert B \rangle_{HS} \equiv
{\mathrm{tr}}(A^\dagger B)$ is a Hilbert space. Therefore
Theorem \ref{prop4}
is an immediate consequence of Proposition \ref{prop3}. $\Box$ \\

The correspondence described in Proposition \ref{prop3} and
Theorem \ref{prop4} has been known
and applied in the quantum optics literature
for some time, see \cite{DAriano00,DAriano01} and
references therein for more details.

In the sequel we always assume without loss of generality that
${\mathtt{H}}_1 = {\mathtt{H}}_2$ and as
in the proof of Proposition \ref{rank1} we denote the trace class
norm of ${\mathfrak{A}}(T)$ by $\tau({\mathfrak{A}}(T))$.
\begin{co} \label{co1}
With the notation from Theorem \ref{prop4}, let $T \in
{\mathtt{HS}}({\mathtt{H}} \otimes {\mathtt{H}})$ be a
Hilbert-Schmidt operator on ${\mathtt{H}} \otimes
{\mathtt{H}}$. Then there exist a family $\{ \lambda_i \}_i$ of
non-negative real numbers,
orthonormal bases $\{ E_i \}_i$ and $\{
F_i \}_i$ of ${\mathtt{HS}}({\mathtt{H}})$ and
${\mathtt{HS}}({\mathtt{H}})$ respectively such that
\begin{equation} \label{rep}
T = \sum_i \lambda_i E_i \otimes F_i. \end{equation}
Moreover we have $\tau({\mathfrak{A}}(T)) = \sum_i \lambda_i$.
\end{co}
\emph{Proof}: This is an immediate consequence of Proposition
\ref{prop3} and Theorem \ref{prop4}. $\Box$ \\

Corollary \ref{co1} can be viewed as an analogue of the Schmidt
decomposition for density operators.

We now apply our results to the separability problem for
density operators on ${\mathtt{H}} \otimes
{\mathtt{H}}$.
It is known that for any operator $A : {\mathtt{HS}}({\mathtt{H}})
\to {\mathtt{HS}}({\mathtt{H}})$
the following identity holds (see
\cite{Schatten70}, page 42,
and also Equation (\ref{tcn}))
\begin{equation} \label{tcn2}
\tau(A) \equiv \inf \left\{ \sum_i \Vert f_i \Vert_2 \, \Vert g_i
\Vert_2 \, \right\vert \left. A = \sum_i \vert f_i \rangle \langle g_i
\vert, \, \vert f_i \rangle, \vert g_i \rangle \in
{\mathtt{HS}}({\mathtt{H}}) \right\}
\end{equation} where the infimum is over all finite decompositions
of $A$ into simple tensors of Hilbert-Schmidt operators.
The next Proposition is
our new necessary separability criterion.
\begin{prop} \label{sepcp}
Let ${\mathtt{H}}$ be a finite dimensional
Hilbert space. Let $\varrho \in {\mathtt{T}}({\mathtt{H}}
\otimes {\mathtt{H}})$ be a density operator. If $\varrho$ is
separable, then
\begin{equation} \label{sepc}
\tau({\mathfrak{A}}(\varrho)) \leq 1.
\end{equation}
\end{prop}
\emph{Proof}: This follows immediately from Equation (\ref{tcn2}) and Theorem
\ref{t1}. $\Box$
\begin{co} \label{p12}
Let ${\mathtt{H}}$ be a finite dimensional
Hilbert space. Let $\vert \psi \rangle \in {\mathtt{H}}
\otimes {\mathtt{H}}$ be a pure state. $\vert \psi \rangle$ is
separable if and only if $\tau({\mathfrak{A}}(P_\psi)) = 1.$
\end{co}
\emph{Proof}: This follows immediately from the proof of
Proposition \ref{rank1}. $\Box$ \\

\begin{rema}
To check whether the separability criterion in Proposition \ref{sepcp}
is satisfied by a given density operator $\varrho$ reduces to the
evaluation of the trace class norm
of the Hilbert-Schmidt operator ${\mathfrak{A}}(\varrho).$
This is completely straightforward using standard linear algebra
packages and accordingly Equation (\ref{sepc}) is
a computable separability criterion for density operators.
\end{rema}

In the next two subsections we compute $\tau({\mathfrak{A}}(\varrho))$
in the situations that $\varrho$ is an isotropic state or a Werner
state respectively. Moreover, in the subsequent subsections we study
other families of states for which Equation (\ref{sepc}) can be
computed.
\subsection{Isotropic states} \label{subsec2}
We continue to use our notation from Section \ref{sec4}. We rewrite
$\varrho_F$ as
\begin{equation} \label{isotrop}
\varrho_F = \frac{1-\alpha_F}{d^2} I + \alpha_F \vert \Psi^+ \rangle
\langle \Psi^+ \vert,
\end{equation} where $\alpha_F \equiv \frac{d^2 F - 1}{d^2 -1}$.
We prove
\begin{prop} \label{iso} Let $F \in [0,d]$ and $\varrho_F$ be the
corresponding isotropic state, then
\[ \tau({\mathfrak{A}}(\varrho_F)) = \left\{
\begin{array}{r@{\quad:\quad}l}
dF & \text{ for } \, \frac{1}{d^2} \leq F \leq 1 \\ \frac{2}{d} - dF
& \mbox{ for } 0 \leq F
< \frac{1}{d^2} \end{array} \right. . \]
\end{prop}
For the proof of this proposition we need Ferrers' formula
\cite{Ferrers55}.
\begin{lem}[Ferrers] \label{Ferrers}
Let $0 < n \in {\mathbb{N}}$ and $a_1, a_2, \ldots, a_n \in
{\mathbb{C}} \backslash \{ 0 \},$ then
\[ \det \left( \begin{array}{ccccc}
1+a_1 & 1 & 1 & \cdots & 1 \\
1 & 1+a_2 & 1 & \cdots & 1 \\
1 & 1 & 1+a_3 & \cdots & 1 \\
\vdots & \vdots & \vdots & \ddots & \vdots \\
1 & 1 & 1 & \cdots & 1+a_n
\end{array} \right) = a_1 a_2 \cdots a_n \left( 1 + \frac{1}{a_1}
+ \cdots + \frac{1}{a_n} \right). \]
\end{lem}
Ferrers' formula follows readily by induction. \\

\noindent\emph{Proof of Proposition \ref{iso}}: Denote by $\{ \vert i \rangle \}_i$
the canonical real basis of ${\mathbb{C}}^d$. From Equation
(\ref{isotrop}) it follows that
\[ {\mathfrak{A}}(\varrho_F) = \frac{\alpha_F}{d} \sum_{ij} \vert E_{ij}
\rangle \langle E_{ij} \vert + \frac{1 - \alpha_F}{d^2} \sum_{ij} \vert
E_{ii} \rangle \langle E_{jj} \vert, \]
where $E_{ij} \equiv \vert i \rangle \langle j \vert$ and where we
use the notation $\vert E_{ij} \rangle$ for $E_{ij}$ to stress
that we think of $E_{ij}$ as an element of ${\mathtt{HS}}({\mathbb{C}}^d).$
Thus ${\mathfrak{A}}(\varrho_F)^\dagger =
{\mathfrak{A}}(\varrho_F)$ and
\begin{eqnarray} {\mathfrak{A}}(\varrho_F)^\dagger
{\mathfrak{A}}(\varrho_F)
& = & \frac{\alpha_F^2}{d^2} \sum_{ij} \vert E_{ij} \rangle \langle E_{ij}
\vert + \frac{1-\alpha_F^2}{d^3} \sum_{ij} \vert E_{ii} \rangle \langle
E_{jj} \vert  \nonumber \\
& = & \frac{\alpha_F^2}{d^2} \sum_{ij \atop i \neq j} \vert E_{ij} \rangle \langle E_{ij}
\vert + \frac{\alpha_F^2}{d^2} \sum_i \vert E_{ii} \rangle \langle
E_{ii} \vert +
\frac{1-\alpha_F^2}{d^3} \sum_{ij} \vert E_{ii} \rangle \langle
E_{jj} \vert. \label{eig} \end{eqnarray}
From the formula (\ref{eig})
we see that $\frac{\alpha_F^2}{d^2}$ is an
Eigenvalue of ${\mathfrak{A}}(\varrho_F)^\dagger
{\mathfrak{A}}(\varrho_F)$ with multiplicity (at least)
$d^2-d$. The second two terms in
Equation (\ref{eig}) act only on the subspace ${\mathcal{S}}_d$ spanned by the
elements $\vert E_{ii} \rangle$. The matrix representation of the
second two terms in Equation (\ref{eig}) in the basis $\{ \vert E_{ii}
\rangle \}_i$ of ${\mathcal{S}}_d$ is
\begin{equation}
\frac{\alpha_F^2}{d^2} \sum_i \vert E_{ii} \rangle \langle
E_{ii} \vert +
\frac{1-\alpha_F^2}{d^3} \sum_{ij} \vert E_{ii} \rangle \langle
E_{jj} \vert \simeq \frac{1 - \alpha_F^2}{d^3}
\left( \begin{array}{ccccc}
1+ \frac{d \alpha_F^2}{1 - \alpha_F^2} & 1 & 1 & \cdots & 1 \\
1 & 1+\frac{d \alpha_F^2}{1 - \alpha_F^2} & 1 & \cdots & 1 \\
1 & 1 & 1+\frac{d \alpha_F^2}{1 - \alpha_F^2} & \cdots & 1 \\
\vdots & \vdots & \vdots & \ddots & \vdots \\
1 & 1 & 1 & \cdots & 1+\frac{d \alpha_F^2}{1 - \alpha_F^2}
\end{array} \right).
\end{equation} The Eigenvalues of this matrix  can readily be
evaluated with the help of Lemma \ref{Ferrers} and are found to be
$\lambda_1 = \frac{\alpha_F^2}{d^2}$ with $d-1$-fold
multiplicity and $\lambda_2 = \frac{1}{d^2}$
(with multiplicity one).
Therefore adding the absolute values of the square roots
of all Eigenvalues, we arrive at
$\tau({\mathfrak{A}}(\varrho_F)) = \vert \alpha_F \vert (d-\frac{1}{d}) +
\frac{1}{d}$. This proves the Proposition. $\Box$
\begin{co}
${\mathrm{tr}}({\mathfrak{A}}(\varrho_F)) = dF$ and
$\Vert \varrho_F \Vert_g = \sqrt{\alpha_F^2 \frac{d^2-1}{d^2}
+ \frac{1}{d^2}}$.
\end{co}
Proposition \ref{iso} implies that an isotropic state $\varrho_F$
is separable if and only if $\tau({\mathfrak{A}}(\varrho_F)) \leq
1$.
\subsection{Werner states} \label{wer2}
\begin{prop} \label{wer} Let $f \in [-1,1]$ and $\varrho_f$ be the
corresponding Werner state, then
\[ \tau({\mathfrak{A}}(\varrho_f)) = \left\{
\begin{array}{r@{\quad:\quad}l}
\frac{2}{d} - f & \text{ for } \, -1 \leq f \leq \frac{1}{d} \\ f &
\mbox{ for } 1 \geq f \geq \frac{1}{d} \end{array} \right. . \]
\end{prop}
\emph{Proof}: We write
\[ {\mathfrak{A}}(\varrho_f) = \frac{d-f}{d^3-d} \sum_{ij} \vert
E_{ii} \rangle \langle E_{jj} \vert + \frac{df-1}{d^3-d} \sum_{ij}
\vert E_{ij} \rangle \langle E_{ji} \vert. \]
An argument as above shows that
${\mathfrak{A}}(\varrho_f)^\dagger{\mathfrak{A}}(\varrho_f)$ has
the simple Eigenvalue $\lambda_0 = \frac{1}{d^2}$ and the degenerate
Eigenvalue $\lambda_1 = \frac{(df-1)^2}{(d^3-d)^2}$ with
multiplicity $d^2 -1.$ This shows that $\tau({\mathfrak{A}}(\varrho_f))
= \frac{\vert df-1 \vert}{d} + \frac{1}{d}.$ $\Box$ \\

Proposition \ref{wer} shows that the criterion in Equation
(\ref{sepc}) is satisfied whenever $f \in \left[ \frac{2}{d}-1, 1
\right]$. This proves that for Werner states the criterion in
Equation (\ref{sepc}) is exact if and only if $d=2$. In higher
dimension $d \geq 3$ there will always be inseparable Werner
states (i.e., those
corresponding to $f \in \left[ \frac{2}{d} -1, 0 \right[$)
which satisfy the criterion in Proposition \ref{sepcp} while
other inseparable Werner states (i.e., those corresponding to
$f \in \left[ -1, \frac{2}{d}-1 \right[$) violate it.
\begin{co}
${\mathrm{tr}}({\mathfrak{A}}(\varrho_f)) = \frac{f+1}{d+1}$ and
$\Vert {\mathfrak{A}}(\varrho_f) \Vert_g = \sqrt{\frac{1+f^2}{d^2-1} -
\frac{2f}{d(d^2-1)}}.$
\end{co}
\subsection{A two qubit example} \label{qubit}
Denote the canonical real basis in ${\mathbb{C}}^2$ by
$\{ \vert 0 \rangle, \vert 1 \rangle \}$ and
consider the following family of states on ${\mathbb{C}}^2 \otimes
{\mathbb{C}}^2$
\[ \varrho_p \equiv p \vert 00 \rangle \langle 00 \vert + (1 -p)
\vert \Phi \rangle \langle \Phi \vert, \]
where $0\leq p \leq 1$ and
$\vert \Phi \rangle = \frac{1}{\sqrt{2}}(\vert 01 \rangle +
\vert 10 \rangle)$. Then
\[  {\mathfrak{A}}(\varrho_p) = p \vert E_{00} \rangle \langle
E_{00} \vert + \frac{1-p}{2} \left( \vert E_{00} \rangle \langle
E_{11} \vert + \vert E_{11} \rangle \langle E_{00} \vert + \vert
E_{10} \rangle \langle E_{01} \vert + \vert E_{01} \rangle \langle
E_{10} \vert \right). \]
It is straightforward to compute the trace class norm of
${\mathfrak{A}}(\varrho_p)$. The result is
\begin{eqnarray*} \tau({\mathfrak{A}}(\varrho_p)) & = &
1 - p + \sqrt{\frac{p^2}{2} + \frac{(1-p)^2}{4} + \frac{p}{2}
\sqrt{p^2 + (1-p)^2}} + \sqrt{\frac{p^2}{2} + \frac{(1-p)^2}{4} -
\frac{p}{2} \sqrt{p^2 + (1-p)^2}} \\
& \geq & 1 -p + p \sqrt{1+ \frac{(1-p)^2}{2p^2}} \geq 1
\end{eqnarray*}
with equality if and only if $p=1$. Therefore Equation
(\ref{sepc}) implies that $\varrho_p$ is separable if and only
if $p = 1$.

\subsection{Bell diagonal states} \label{Bell}
We continue to use our notation from Section \ref{Bellgamma} but assume
now that (without loss of generality) $\{ \vert 1 \rangle, \vert 2 \rangle
\}$ denotes the canonical real basis in ${\mathbb{C}}^2$.
Let $\varrho$ be a Bell diagonal state, i.e.,
a density operator on ${\mathbb{C}}^2 \otimes
{\mathbb{C}}^2$ of the form
\[ \varrho =\sum_{i=0}^3 \lambda_i \vert \Psi_i \rangle \langle
\Psi_i \vert. \]
Then \begin{eqnarray*} {\mathfrak{A}}(\varrho) & = &
\frac{\lambda_0 + \lambda_3}{2}
\left( \vert E_{11} \rangle \langle E_{11} \vert + \vert E_{22}
\rangle \langle E_{22} \vert \right)
+ \frac{\lambda_1 + \lambda_2}{2}
\left( \vert E_{11} \rangle \langle E_{22} \vert + \vert E_{22}
\rangle \langle E_{11} \vert \right) \\ & & +
\frac{\lambda_0 - \lambda_3}{2}
\left(\vert E_{12} \rangle \langle E_{12} \vert +
\vert E_{21} \rangle \langle E_{21} \vert \right)
+ \frac{\lambda_1 - \lambda_2}{2}
\left( \vert E_{12} \rangle \langle E_{21} \vert +
\vert E_{21} \rangle \langle E_{12} \vert \right).
\end{eqnarray*}
Hence
\begin{eqnarray*} {\mathfrak{A}}(\varrho)^\dagger
{\mathfrak{A}}(\varrho) & = &
\frac{(\lambda_0 + \lambda_3)^2 + (\lambda_1 + \lambda_2)^2}{4}
\left( \vert E_{11} \rangle \langle E_{11} \vert + \vert E_{22}
\rangle \langle E_{22} \vert \right)  \\ & &
+ \frac{(\lambda_0 + \lambda_3) (\lambda_1 + \lambda_2)}{2}
\left( \vert E_{11} \rangle \langle E_{22} \vert + \vert E_{22}
\rangle \langle E_{11} \vert \right) \\ & & +
\frac{(\lambda_0 - \lambda_3)^2 +
(\lambda_1 - \lambda_2)^2}{4}
\left(\vert E_{12} \rangle \langle E_{12} \vert +
\vert E_{21} \rangle \langle E_{21} \vert \right) \\ & &
+ \frac{(\lambda_1 - \lambda_2) (\lambda_0 - \lambda_3)}{2}
\left( \vert E_{12} \rangle \langle E_{21} \vert +
\vert E_{21} \rangle \langle E_{12} \vert \right).
\end{eqnarray*}
It is straightforward to compute the trace class norm of
${\mathfrak{A}}(\varrho)$. The result is \alpheqn
\begin{eqnarray}
\tau({\mathfrak{A}}(\varrho)) & = &
\frac{1}{2} \left( 1+ \vert \lambda_0 +
\lambda_3 - \lambda_1 - \lambda_2 \vert +
\vert \lambda_1 - \lambda_2 \vert + \vert \lambda_0 - \lambda_3
\vert + \left\vert \, \vert \lambda_0 - \lambda_3 \vert - \vert
\lambda_1 - \lambda_2 \vert \, \right\vert \right) \\
\label{1234} & = & \left\{
\begin{array}{l@{\quad:\quad}l}
2 \max_i \lambda_i & \text{ if } \, \left\{
\begin{array}{r@{}l} \text{either} \, &
\lambda_0 + \lambda_3 \geq \lambda_1 +
\lambda_2, \, \, \, \, \vert \lambda_0 - \lambda_3 \vert \geq
\vert \lambda_1 - \lambda_2 \vert \\
\text{or} \, \, \, \, & \lambda_0 + \lambda_3 \leq \lambda_1 +
\lambda_2, \, \vert \lambda_0 - \lambda_3 \vert \leq
\vert \lambda_1 - \lambda_2 \vert \end{array} \right.
\\ 1 - 2 \min_i \lambda_i & \mbox{ otherwise}. \end{array} \right. .
\end{eqnarray}
\reseteqn
To see Equation (\ref{1234}) note that \begin{itemize}
\item if $\lambda_0 + \lambda_3 \geq
\lambda_1 + \lambda_2$ and $\vert \lambda_0 - \lambda_3 \vert \geq
\vert \lambda_1 - \lambda_2 \vert$, then $\max_i \lambda_i = \max \{
\lambda_0, \lambda_3 \}$. Similarly, if $\lambda_1 + \lambda_2 \geq
\lambda_0 + \lambda_3$ and $\vert \lambda_1 - \lambda_2 \vert \geq
\vert \lambda_0 - \lambda_3 \vert$, then $\max_i \lambda_i = \max \{
\lambda_1, \lambda_2 \}$. \item Conversely, if $\lambda_0 + \lambda_3
\geq \lambda_1 + \lambda_2$ and $\vert \lambda_0 - \lambda_3 \vert
< \vert \lambda_1 - \lambda_2 \vert$, we find $\min_i \lambda_i =
\min \{ \lambda_1, \lambda_2 \}$. Similarly, if $\lambda_1 + \lambda_2
\geq \lambda_0 + \lambda_3$ and $\vert \lambda_1 - \lambda_2 \vert
< \vert \lambda_0 - \lambda_3 \vert$, we find $\min_i \lambda_i =
\min \{ \lambda_0, \lambda_3 \}$. \item
Note also that if $\max_i \lambda_i \geq \frac{1}{2},$ then
we have either the situation that
$\lambda_0 + \lambda_3 \geq \lambda_1 + \lambda_2$ and
$\vert \lambda_0 - \lambda_3 \vert \geq
\vert \lambda_1 - \lambda_2 \vert$ or that
$\lambda_0 + \lambda_3 \leq \lambda_1 + \lambda_2$ and
$\vert \lambda_0 - \lambda_3 \vert \leq
\vert \lambda_1 - \lambda_2 \vert$. \\
\emph{Proof}:
To see this, assume without loss of generality
that $\lambda_0 = \max_i \lambda_i = \frac{1}{2} +\delta$ for some
$\delta
\geq 0.$ Write $\lambda_3 = \lambda_0 - \epsilon$, for $\epsilon \geq 0.$
Now assume that $\lambda_0 + \lambda_3 \geq \lambda_1 + \lambda_2$, but
$\vert \lambda_0 - \lambda_3 \vert < \vert \lambda_1 - \lambda_2 \vert$.
Then $\epsilon = \lambda_0 - \lambda_3 < \vert \lambda_1 - \lambda_2 \vert
\leq \lambda_1 + \lambda_2 = 1 - \lambda_0 - \lambda_3 = - 2 \delta +
\epsilon$. Hence $-2 \delta > 0$. This is a contradiction.
$\Box$
\item $\vert \lambda_0 - \lambda_3 \vert = \vert \lambda_1 - \lambda_2
\vert$ implies $2\max_i \lambda_i = 1 - 2 \min_i \lambda_i$.
\item $\lambda_0 + \lambda_3 = \lambda_1 + \lambda_2$ implies
$2\max_i \lambda_i = 1 - 2 \min_i \lambda_i$.
\end{itemize}
Thus
$\max_i \lambda_i \geq \frac{1}{2}$ implies that
$\tau({\mathfrak{A}}(\varrho)) = 2 \max_i \lambda_i$. Therefore we
conclude that $\tau({\mathfrak{A}}(\varrho)) \leq 1$ if and only
if $\varrho$ is separable. \\

\noindent\emph{Conclusion of the proof of Theorem
\ref{gammabell}}: The remaining inequality in the proof of Theorem
\ref{gammabell} now follows immediately from $\tau({\mathfrak{A}}(\varrho))
\leq \Vert \varrho \Vert_\gamma$ and the results of this
subsection. $\Box$
\subsection{A two qutrit example} \label{qutrit}
Consider ${\mathbb{C}}^3 \otimes {\mathbb{C}}^3$ and let
$\{ \vert 0 \rangle, \vert 1 \rangle, \vert 2 \rangle \}$ be
the canonical real basis in ${\mathbb{C}}^3$. Consider
the following family of qutrit mixed states defined on
${\mathbb{C}}^3 \otimes {\mathbb{C}}^3$
\begin{equation}
\varrho_\alpha := \frac{2}{7} \left\vert \Psi^+_{(3)} \right\rangle
\left\langle
\Psi^+_{(3)} \right\vert + \frac{\alpha}{7} \sigma_+ +\frac{5-\alpha}{7}
\sigma_-, \end{equation}
where we restrict ourselves to the parameter range
$2 \leq \alpha \leq 5$, and where
\begin{eqnarray*} \left\vert \Psi^+_{(3)} \right\rangle & \equiv &
\frac{1}{\sqrt{3}} \left( \vert
0 \rangle \vert 0 \rangle + \vert 1 \rangle \vert 1 \rangle +
\vert 2 \rangle \vert 2 \rangle
\right), \\ \sigma_+ & \equiv & \frac{1}{3} \left( \vert 0 \rangle
\vert 1 \rangle \langle 0 \vert \langle 1 \vert + \vert 1 \rangle
\vert 2 \rangle \langle 1 \vert \langle 2 \vert + \vert 2 \rangle
\vert 0 \rangle \langle 2 \vert \langle 0 \vert \right) \\
\sigma_- & \equiv & \frac{1}{3} \left( \vert 1 \rangle
\vert 0 \rangle \langle 1 \vert \langle 0 \vert + \vert 2 \rangle
\vert 1 \rangle \langle 2 \vert \langle 1 \vert + \vert 0 \rangle
\vert 2 \rangle \langle 0 \vert \langle 2 \vert \right). \end{eqnarray*}
It is known \cite{Horodecki98} that $\varrho_\alpha$ is (i) separable
if and only if $2 \leq \alpha \leq 3$, (ii) bound entangled if and
only if $3 < \alpha \leq 4$ and (iii) entangled and distillable if
and only if $4 < \alpha \leq 5$.

We have
\begin{eqnarray*}
{\mathfrak{A}}(\varrho_\alpha) & = & \frac{2}{21} \left(\vert
E_{00} \rangle \langle E_{00} \vert + \vert E_{01} \rangle \langle
E_{01} \vert + \vert E_{02} \rangle \langle
E_{02} \vert + \vert E_{10} \rangle \langle
E_{10} \vert + \vert E_{11} \rangle \langle
E_{11} \vert \right. \\ & & \, \, \, \, \, \, \, \, \, \, \, + \left.
\vert E_{12} \rangle \langle
E_{12} \vert + \vert E_{20} \rangle \langle
E_{20} \vert + \vert E_{21} \rangle \langle
E_{21} \vert + \vert E_{22} \rangle \langle
E_{22} \vert \right) \\ & & + \frac{\alpha}{21} \left(
\vert E_{00} \rangle \langle
E_{11} \vert + \vert E_{11} \rangle \langle
E_{22} \vert + \vert E_{22} \rangle \langle
E_{00} \vert \right) \\ & & + \frac{5-\alpha}{21} \left(
\vert E_{11} \rangle \langle
E_{00} \vert + \vert E_{22} \rangle \langle
E_{11} \vert + \vert E_{00} \rangle \langle
E_{22} \vert \right).
\end{eqnarray*}
Accordingly
\begin{eqnarray*}
{\mathfrak{A}}(\varrho_\alpha)^\dagger
{\mathfrak{A}}(\varrho_\alpha) & = & \frac{4}{441} \left(
\vert E_{01} \rangle \langle E_{01} \vert +
\vert E_{02} \rangle \langle E_{02} \vert +
\vert E_{10} \rangle \langle E_{10} \vert +
\vert E_{20} \rangle \langle E_{20} \vert +
\vert E_{12} \rangle \langle E_{12} \vert +
\vert E_{21} \rangle \langle E_{21} \vert \right) \\ & & \, \, \, \,
+ \frac{2 \alpha^2 - 10 \alpha +29}{441} \left(
\vert E_{00} \rangle \langle E_{00} \vert +
\vert E_{11} \rangle \langle E_{11} \vert +
\vert E_{22} \rangle \langle E_{22} \vert \right) \\ & & \, \, \, \,
+ \frac{10 + 5 \alpha - \alpha^2}{441} \left(
\vert E_{00} \rangle \langle E_{11} \vert +
\vert E_{00} \rangle \langle E_{22} \vert +
\vert E_{11} \rangle \langle E_{00} \vert +
\vert E_{11} \rangle \langle E_{22} \vert \right. \\ & & \, \, \, \,
+ \left.
\, \, \, \,
\vert E_{22} \rangle \langle E_{00} \vert +
\vert E_{22} \rangle \langle E_{11} \vert \right).
\end{eqnarray*}
The corresponding Eigenvalue problem can readily be solved using
Ferrers' formula and we arrive at
\begin{equation} \label{tauqutrit}
\tau({\mathfrak{A}}(\varrho_\alpha)) = \frac{19}{21} + \frac{2}{21}
\sqrt{19 - 15 \alpha + 3 \alpha^2}.
\end{equation}
It is easy to see that $\tau({\mathfrak{A}}(\varrho_\alpha)) \leq
1$ if and only if $2 \leq \alpha \leq 3$, i.e., if and only if
$\varrho_\alpha$ is separable. This example shows that there are
inseparable (bound entangled)
states which violate Equation (\ref{sepc}) but satisfy
the reduction criterion. \\
\subsection{Concluding remarks}
It is known that for $d \geq 3$ all inseparable
Werner states violate the Peres-Horodecki positive partial
transpose (ppt) criterion for separability
(see \cite{Peres96,Horodecki96b})
but do not violate the reduction
criterion introduced in \cite{Horodecki97}. As moreover
the bound entangled states
in subsection \ref{qutrit} satisfy the ppt criterion, it
follows from our results in subsections \ref{wer2}
and \ref{qutrit} that the
separability criterion in Equation (\ref{sepc}) is neither
stronger nor weaker than
the positive partial transpose criterion.
Moreover, it also follows that the
criterion Equation (\ref{sepc}) is
not weaker than the reduction
criterion for separability. By the results of
\cite{VollbrechtW02} this
also implies that our criterion is not weaker than the entropic
separability criteria based on the generalized R\'enyi and Tsallis
entropies. The example in subsection \ref{qubit}
implies that the separability criterion in
Equation (\ref{sepc}) is also not weaker than
the criterion proposed by Nielsen and
Kempe in \cite{NielsenK01} (as the criterion in \cite{NielsenK01}
completely characterizes the separability properties
of isotropic states in arbitrary dimension, but
fails for the states $\varrho_p$ discussed in subsection
\ref{qubit} and for all inseparable Werner states
in dimension $d \geq 3$, see also \cite{VollbrechtW02}).
Finally, violating our criterion does not imply distillability.
\\

\noindent\textbf{Acknowledgement} Thanks to the members of
the Quantum Optics \& Information Group at Pavia for their hospitality and
in particular to Giacomo Mauro D'Ariano and Shashank Virmani
for helpful
discussions about entanglement and about
quantum information in general.
Funding by the European Union project ATESIT (contract
IST-2000-29681) is gratefully acknowledged.

\end{document}